\documentclass[article]{elsarticle}

\usepackage{lineno,hyperref}
\usepackage{amsfonts}
\usepackage{amsmath}
\usepackage{breqn}
\modulolinenumbers[5]

\journal{New Astronomy}

\begin{document}

\begin{frontmatter}

\title{Hourglass Magnetic Field from a Survey of Current Density Profiles}


\author[mysecondaryaddress]{Gianfranco Bino}

\author[mysecondaryaddress,mymainaddress]{Shantanu Basu}

\author[mysecondaryaddress]{Mahmoud Sharkawi}

\author[mysecondaryaddress,mymainaddress]{Indrani Das}

\address[mysecondaryaddress]{Department of Applied Mathematics, University of Western Ontario, London, ON, N6A 5B7, Canada.}
\address[mymainaddress]{Department of Physics \& Astronomy, University of Western Ontario, London, ON, N6A 3K7, Canada.}

\begin{abstract}
Modeling the magnetic field in prestellar cores can serve as a useful tool for studying the initial conditions of star formation. The analytic hourglass model of \cite{ewertowski2013mathematical} provides a means to fit observed polarimetry measurements and extract useful information. The original model does not specify any radial distribution of the electric current density. Here, we perform a survey of possible centrally-peaked radial distributions of the current density, and numerically derive the full hourglass patterns. Since the vertical distribution is also specified in the original model, we can study the effect of different ratios
of vertical to radial scale length on the overall hourglass pattern. Different values of this ratio may correspond to different formation scenarios for prestellar cores.
We demonstrate the flexibility of our model and how it can be applied to a variety of magnetic field patterns.
\end{abstract}

\begin{keyword}
stars: formation --- ISM: magnetic fields --- mathematical model
\end{keyword}

\end{frontmatter}

%\linenumbers

\section{Introduction}
\label{introduction}
Magnetic fields are thought to play an important role in the formation of stars, dating back to the ideas of \citet{mestel1956}, and described in many subsequent reviews \citep[e.g.,][]{mouschovias78,shu87,shu99,mou1999,wurster18}. Under certain circumstances, the morphology of the magnetic field takes the form of an hourglass shape. When the magnetic force cannot balance gravity, the field lines are dragged inwards and pinched at the center as a result of gravitational contraction. This occurs when the flux is frozen-in with the fluid. An hourglass pattern can also occur with partial flux-freezing (e.g., with a neutral-ion drift) but with a less pinched hourglass configuration than in the perfectly flux-frozen case \citep[see][]{bas09}. 

\citet{ewertowski2013mathematical} derived a mathematical model that is valid for arbitrary radial distributions of electric current density. The model was first applied to simulation data, but later deployed by \citet{binobasu2021} to modeling the polarization state and magnetic properties of the prestellar core FeSt 1--457, for which near-infrared polarimetry had been obtained by \citet{Kandori_2017}. The directional polarization segments of FeSt 1--457 had been fit \citep{Kandori_2017, kan3} with a model of parametric parabolic curves.

Although this method demonstrated good fits, it provided little direct physical information about the actual magnetic field. An estimate of the mean magnetic field strength required a measurement of the fluctuations of polarization direction relative to the mean values, and use of the Davis-Chandrasekhar-Fermi (DCF) method \citep{davis1951,CF1953}.  A more complete approach is to fit a physical model of the magnetic field, as done by \cite{binobasu2021}, where the model then can yield values of the magnetic field components 
%$B_r$ and $B_z$ 
at all locations within the core. 
%This facilitated a detailed discussion on the mass-to-flux ratio governing the core at different instances in its evolution along with inference on potential contraction mechanisms governing the collapse. 
In this paper, we extend the use of the model developed by \citet{ewertowski2013mathematical} by directly modeling the radial distribution of the current density in addition to its vertical distribution. The current distribution is assumed to be separable in this formulation. \citet{ewertowski2013mathematical} adopted a Gaussian distribution of current density in the vertical direction and found an analytic solution for the magnetic field in series form, in which each coefficient in the series depends upon an integral that contains the radial current density profile.
They demonstrated how one could determine least-squares best fit values of the coefficients by fitting the model to magnetodhydrodynamic (MHD) simulation data. In the formulation of \citet{ewertowski2013mathematical}, the radial current density is not determined explicitly, but is defined implicitly by the value of the fitted coefficients. Here, we instead specify distributions of the radial current density and determine the resulting hourglass magnetic field pattern. A fitting approach using this model would fit the parameters of the current density model in order to find the best match with either a simulation result or observed polarization segments. In the latter case, a more complete comparison would be to compare the polarization segments with a synthetic polarization map that is generated from the magnetic field model along with the cohabiting gas and dust content \citep[e.g.,][]{tomisaka11,kataoka2012,binobasu2021}.

There is a rapidly-growing literature of polarimetry measurements of star-forming regions, utilizing an array of instruments like the {\it Planck} satellite, the {\it James Clerk Maxwell Telescope} (JCMT), the {\it Atacama Large Millimeter/submillimeter Array} (ALMA), and the {\it Stratospheric Observatory for Infrared Astronomy} (SOFIA). Hence, analytic magnetic field models hold great promise for future analysis and interpretation. On Galactic scales and within molecular clouds, \citet{planck2015,planck2016} have shown that there is a large-scale ordering of the magnetic field, and that the magnetic field generally dominates the energy densities of turbulence and self-gravity. Furthermore, very dense structures in molecular clouds seem to be aligned perpendicular to the magnetic field, presumably because their self-gravity becomes important and contraction can occur preferentially along the large-scale magnetic field direction.
Once a large-scale ordering of the magnetic field is determined, the fluctuations of the polarization vectors relative to the mean can be used to infer a mean magnetic field strength (plane-of-sky value) using the DCF method that is based on a model of Alfv\'enic fluctuations. 
%CAN INSERT A SENTENCE HERE ABOUT PLANCK B FIELD DETERMINATION
\citet{Pattle2017} used results obtained with JCMT and employed the DCF method to determine the strength of the magnetic field on the large scales of the OMC--1 region of the Orion A filament. They found that the magnetic energy density was slightly greater than the self-gravitational energy, measured through a normalized mass-to-flux ratio $\mu \equiv 2 \pi G^{1/2} \Sigma/B$\footnote{Here $\Sigma$ is the mass column density and $B$ is the magnetic field strength.} that was estimated to be $\sim 0.4$.  On smaller scales of dense star-forming cores, where gravity is the dominant force, the magnetic field can be dragged inward to make an hourglass-like morphology. This is now observed in several star-forming regions including NGC 1333 IRAS 4A \citep{Giratetal2006,frau2011}, OMC--1 \citep{Schleuning1998}, L1448 IRS 2 \citep{kwon2019} and G31.41+0.31 \citep{Giratetal2009}.

The plan of this paper is as follows. In Section \ref{Methods} we describe our methodology, and in Section \ref{results} we present our numerical results. Sections \ref{discussion} and \ref{conclusion} are used to discuss and summarize our findings, respectively. 

\section{Methods}
\label{Methods}
%\subsection{Methodology and Techniques}
In the \citet{ewertowski2013mathematical} model, a purely poloidal magnetic field is written as $\boldsymbol{B}=\boldsymbol{B}_d + \boldsymbol{B}_0$, where $\boldsymbol{B}_d= B_r \boldsymbol{\hat{r}} + B_z \boldsymbol{\hat{z}}$ and $\boldsymbol{B}_0=B_0\boldsymbol{\hat{z}}$. The boundary conditions are imposed in a way such that $\boldsymbol{B}$ must attain the background value $\boldsymbol{B}_0$ as $z \rightarrow \pm \infty$, and for $r\geq R$, where $R$ is the core radius. The induced field in the star-forming region, $\boldsymbol{B}_d$, is obtained from a magnetic vector potential $\boldsymbol{A} = A(r,z) \boldsymbol{\hat{\varphi}}$, where
\begin{equation}
A(r,z) = \frac{4 \pi}{c} \int_{-\infty}^{\infty} \int_{0}^{R} G(r,z,\xi, \eta) j(\xi, \eta ) \xi d\xi d \eta \,.
\end{equation}
The source term is the current density $j(r,z)$ and
\begin{equation}
G(r,z,\xi, \eta) =  \sum_{m=1}^{\infty} \frac{J_1(\sqrt{\lambda_m} \xi) J_1(\sqrt{\lambda_m} r)e^{-\sqrt{\lambda_m}|z-\eta|}}{R^2 \sqrt{\lambda_m}\, [J_2(\sqrt{\lambda_m} R)]^2} 
\label{eq:GreenFunc}
\end{equation}
is the Green's function solution of the underlying partial differential equation for $A(r,z)$ \citep[see][for details]{ewertowski2013mathematical}. Furthermore,
\begin{equation}
\lambda_m = \left(\frac{a_{m,1}}{R}\right)^2, \quad m \in \mathbb{N},
\label{eq:lambdam}    
\end{equation}
where $a_{m,1}$ is the $m^{th}$ positive root of $J_1 (x)$ and $a_{m,1} < a_{m+1,1}$. The notation $J_n$ denotes Bessel functions of the first kind of order $n$.
If we assume that the current density has a separable form
\begin{equation}
j(r,z) \equiv f(r)g(z),
\label{eq:current}
\end{equation}
and that $g(z) =e^{-z^2 / h^2}$, \citet{ewertowski2013mathematical} showed that
\begin{equation}
A(r,z) = \frac{2}{h \sqrt{\pi}}\sum_{m=1}^{\infty} k_m e^{-h^2 \lambda_m / 4} J_1(\sqrt{\lambda_m} r) \int_{-\infty}^{\infty} g(\eta) e^{-\sqrt{\lambda_m}|z-\eta|} d \eta ,
\label{eq:3}
\end{equation}
where
\begin{equation}
k_m = \frac{2h \pi^{3/2} e^{h^2 \lambda_m / 4}}{cR^2 \sqrt{\lambda_m}[J_2(\sqrt{\lambda_m} R)]^2} \int_{0}^{R} f(\xi)  J_1(\sqrt{\lambda_m} \xi) \xi d \xi ,
\label{eq:4}
\end{equation}
and $c$ is the speed of light. 
The integral in equation (\ref{eq:3}) can be evaluated analytically for the adopted Gaussian profile of $g(z)$, leading to 
\begin{dmath}
\label{eq:A_series2}
	A(r,z) =\sum_{m=1}^{\infty}  k_m J_1(\sqrt{\lambda_m}  r) \left[ \mathrm{erfc} \left( \frac{\sqrt{\lambda_m} h}{2} + \frac{z}{h} \right)e^{\sqrt{\lambda_m} z}  + \mathrm{erfc} \left( \frac{\sqrt{\lambda_m} h}{2} - \frac{z}{h} \right)e^{-\sqrt{\lambda_m} z} \right] \,.
\end{dmath}
The magnetic field components, including the local and background contribution, are 
% Bz in terms of A
\begin{align}
	B_z &= \frac{1}{r}\frac{\partial{(A \ r)} }{\partial{r}} + B_0 \label{eq:Bz}  \, , \\
	B_r &= -\frac{\partial{A} }{\partial{z}}  \, . \label{eq:Br}
\end{align}
These yield the explicit solutions
% Br

\begin{dmath}
 B_r =\sum_{m=1}^{\infty}  k_m \sqrt{\lambda_m }J_1(\sqrt{\lambda_m}  r) \left[ \mathrm{erfc} \left( \frac{\sqrt{\lambda_m} h}{2} - \frac{z}{h} \right) e^{-\sqrt{\lambda_m} z}   - \mathrm{erfc} \left( \frac{\sqrt{\lambda_m} h}{2} + \frac{z}{h} \right)e^{\sqrt{\lambda_m} z} \right] \, , \\
 \label{eq:Br2}
\end{dmath} 
    
% Bz
\begin{dmath}
B_z =\sum_{m=1}^{\infty}  k_m \sqrt{\lambda_m}J_0(\sqrt{\lambda_m}  r) \left[ \mathrm{erfc} \left( \frac{\sqrt{\lambda_m} h}{2} + \frac{z}{h} \right)e^{\sqrt{\lambda_m} z}  \\ + \mathrm{erfc} \left( \frac{\sqrt{\lambda_m} h}{2} - \frac{z}{h} \right)e^{-\sqrt{\lambda_m} z} \right] + B_0\,.\label{eq:Bz2}
\end{dmath}

\medskip

valid for $(r,z) \in D$. Note that for $r \geq R$, the magnetic field is equal to the background value $\boldsymbol{B}_0$. 

\begin{figure}
    \centering
    \includegraphics[width=0.55\textwidth]{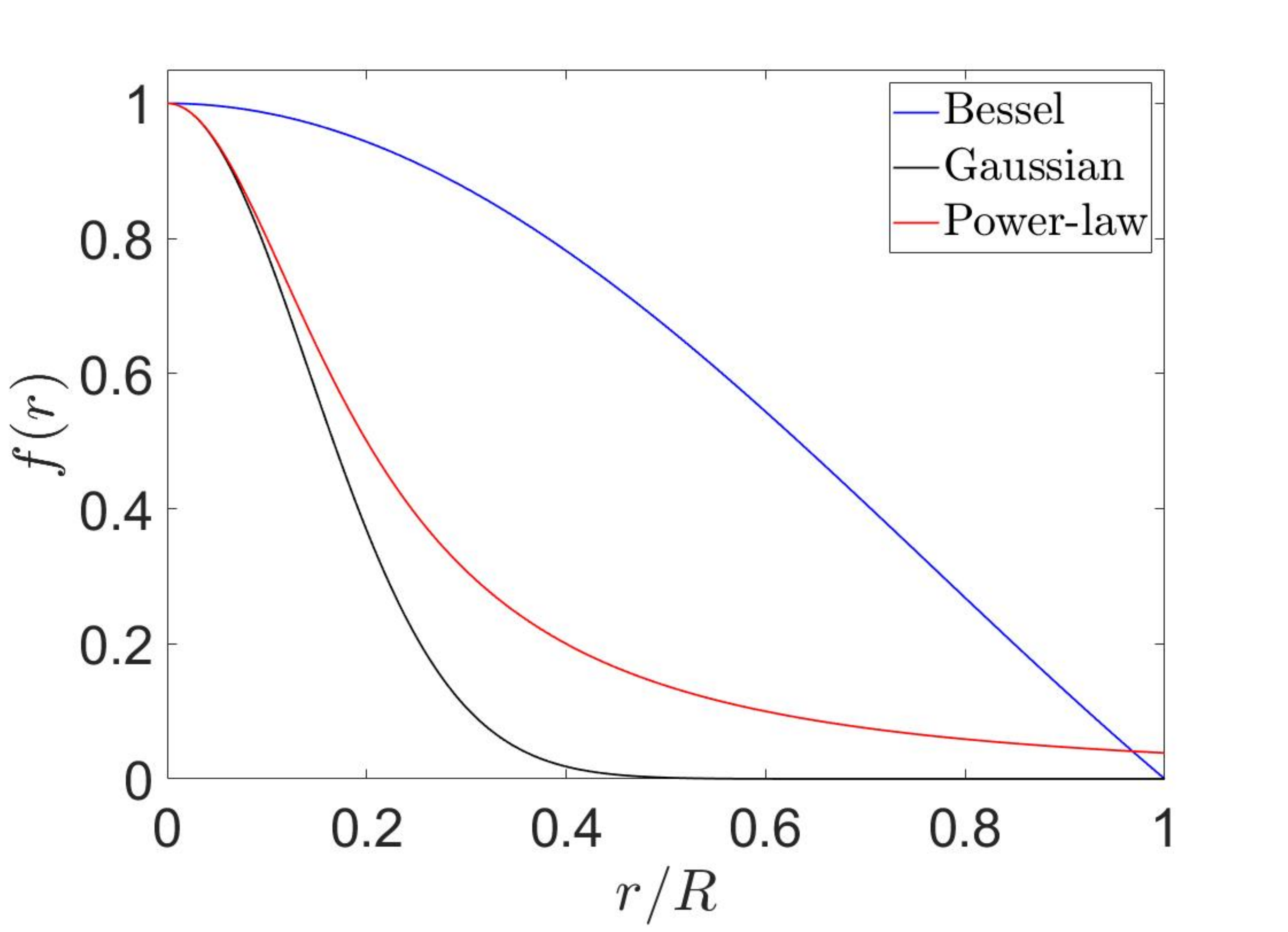}
    \caption{Radial profiles from equation (\ref{eq:gaussian_bessel}). For the Gaussian and power-law profiles we have adopted $l = 0.2R$. }
    \label{fig:r_profiles}
\end{figure}

In \citet{ewertowski2013mathematical}, the coefficients $k_m$ were evaluated to fit the data. Here, we instead evaluate the integral in equation (\ref{eq:4}) directly. Therefore, we must provide a model for the current distribution along the $r$-direction. 
%We furthermore make the assumption that the distributions along each positional direction are independent of one another. 
We survey various distributions that take on the following functional forms:
\begin{equation}
  f(r) =
    \begin{cases}
      e^{-r^2 / l^2} & \text{Gaussian,}\\
      J_0 \left( \alpha_{0,1} r / R \right) & \text{Bessel,}\\
      1 / (1 + r^2/l^2) & \text{power law.}
     \end{cases}  
\label{eq:gaussian_bessel}     
\end{equation}
Here $\alpha_{0,1}$ is the first root of $J_0$, and $l$ is the radial scale length that is treated as a free parameter. These functions represent a range of centrally-peaked distributions. Examples of these profiles are illustrated in Figure \ref{fig:r_profiles}. Evaluating the coefficients $k_m$ from equation (\ref{eq:4}) involves a definite integral that can be solved numerically using various quadrature techniques. 

\section{Results}
\label{results}
%\subsection{Current Density Models}
\begin{figure*}
\centering
  \includegraphics[width=1\textwidth]{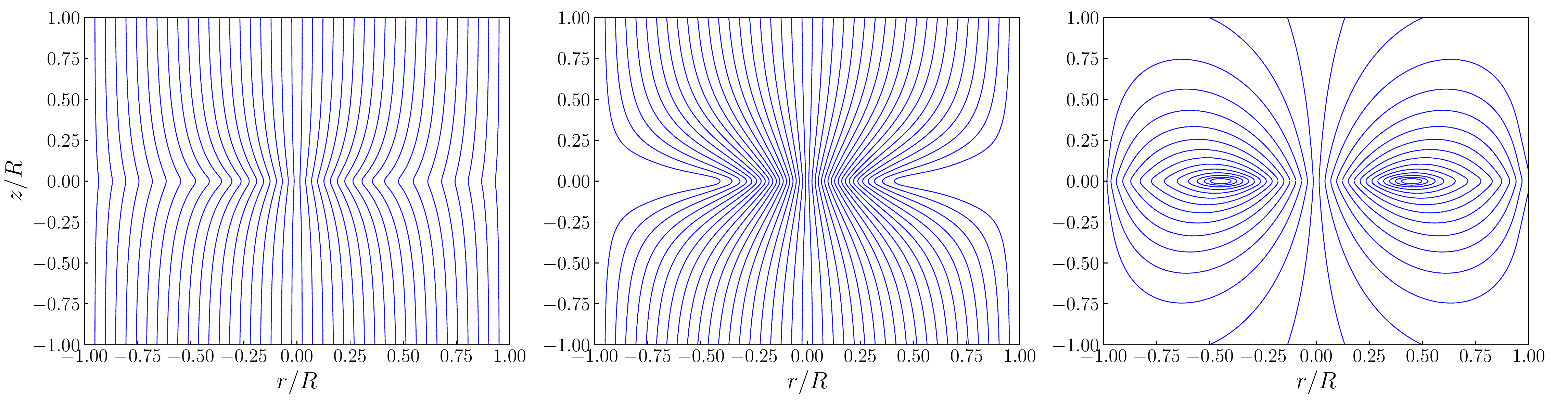}
  \vspace{-0.8cm}
  \caption{Magnetic field line morphology for model 1 (Gaussian radial and Gaussian vertical distributions) demonstrated for increasing central-to-background field ratio $B_c/B_0 \in [4.36, 16.16, 1475.56]$  (from left to right), with fixed ratio of vertical and radial scale length, $h/l = 0.02$.}
  \label{model_1a}
\end{figure*}
\begin{figure*}
\centering
\hspace{-0.6cm}
  \includegraphics[width=1\textwidth]{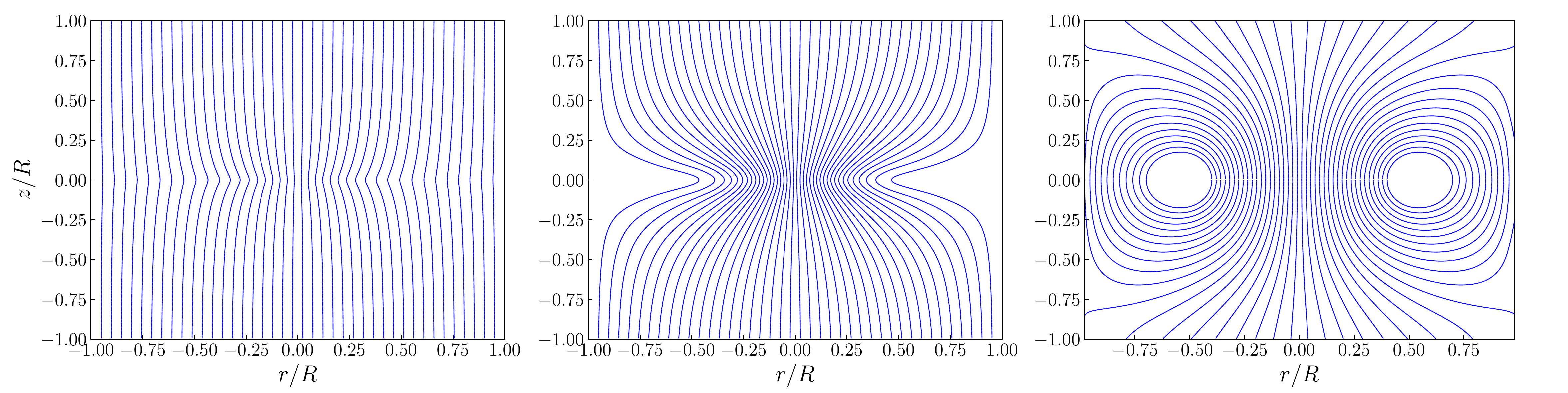}
  \vspace{-0.3cm}
  \caption{Magnetic field morphology for model 1 (Gaussian radial and Gaussian vertical distributions) demonstrated for the scale length ratios $h/l \in [0.01, 0.1, 0.75]$ (from left to right) with respective $B_c/B_0 \in [2.63, 13.69, 51.95]$.}
 \label{model_1b}
\vspace{0.5cm}
\includegraphics[width=1\textwidth]{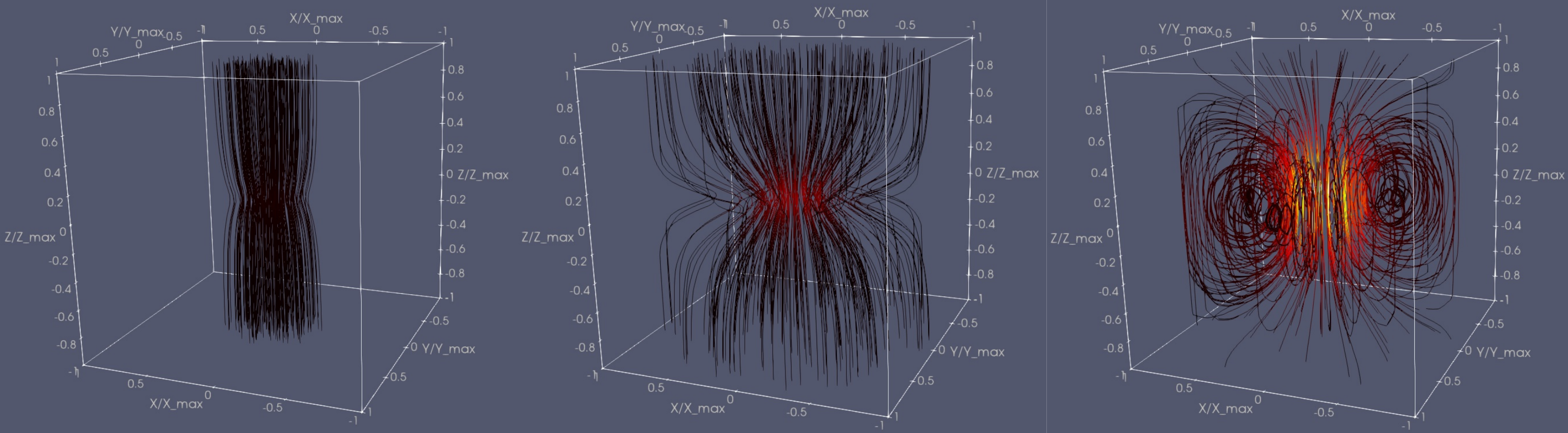}
  \caption{Three-dimensional renderings of the magnetic field morphology in Figure \ref{model_1b}. The color gradient is representative of the field strength ranging from the black being weakest to yellow being strongest.}
  %\caption{Three-dimensional renderings of the magnetic field morphology in Figure \ref{model_1b}. The color gradient is representative of the field strength ranging from the black being weakest to yellow being strongest.}
  \label{model_1_3d}
\end{figure*}

\begin{figure*}[ht]
\centering
\hspace{-0.6cm}
  \includegraphics[width=1\textwidth]{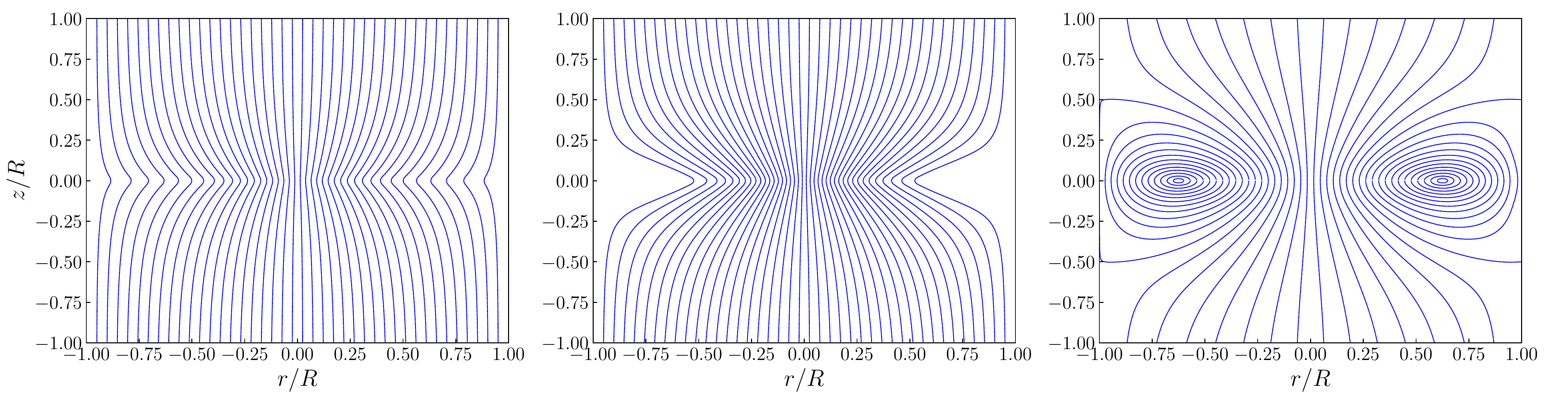}
    \vspace{-0.3cm}
    \caption{Magnetic field morphology for model 2 (Bessel radial and Gaussian vertical distributions) demonstrated for increasing vertical scale length $h/R \in [0.01, 0.025, 0.1]$ (from left to right) with respective $B_c/B_0 \in [5.07, 10.53, 32.71]$. The characteristic radial scale length for the Bessel function in this model is taken as the core radius itself and remains unchanged.}
 \label{model_2}
 \vspace{0.5cm}
  \includegraphics[width=1\textwidth]{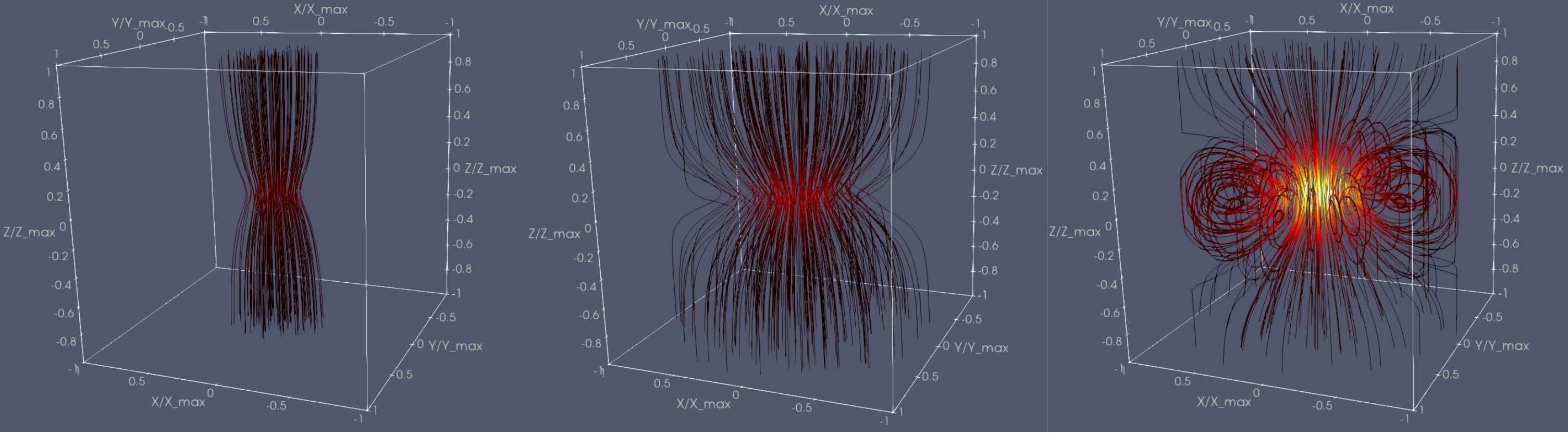}
  \caption{Three-dimensional renderings of the magnetic field morphology in Figure \ref{model_2}. The color gradient is representative of the field strength ranging from the black being weakest to yellow being strongest.}
  \label{model_2_3d}
\end{figure*}

 
We study combinations of current density distributions using the functional forms introduced earlier. Fixing the vertical function as $g(z) =e^{-z^2 / h^2}$, we construct our models by using the adopted forms of $f(r)$, where $l$ and $h$ represent the scale lengths along the $r$- and $z$- directions, respectively. 
The model with a Gaussian distribution along the radial direction is called model 1. Similarly, model 2 has a Bessel function for $f(r)$. Lastly, we define model 3 to have $f(r)$ as a power-law distribution.
The model 1 Gaussian radial profile is more centrally concentrated than the model 3 power-law radial profile for a given scale length $l$. 
%and 3 provide similar distributions for the current over the $r$- direction with model 1 concentrating the current about $r=0$ slightly more. 
Model 2 provides a current density that tapers off less aggressively toward zero than models 1 or 3 as $r \rightarrow R$. The characteristic scale length for the Bessel function in model 2 is the core radius itself and remains unchanged. Physically, this differs from models 1 and 3 in a shallower convergence of the core generated field to the background field as $r \rightarrow R$. The inputs to our numerical models are the core radius $R$, the radial scale length $l$, the vertical scale length $h$, and the background magnetic field $B_0$.
We investigate the change in the field morphology by varying $B_0$, $l$, and $h$. 
We truncate the infinite sum in equation (\ref{eq:3}) at the $m = 3$ term. The higher order terms are representative of higher frequency components of the magnetic waveform given by the Bessel series.  In Figure \ref{model_1a} we show model 1 and the effect of varying the background magnetic field, with $B_c/B_0 \in [4.36, 16.16, 1475.56]$, in which $B_c$ is the central magnetic field strength. We evaluate the vector of coefficients $\boldsymbol{k} = [k_1, k_2,k_3]$ using the trapezoidal rule for integration and find that
\begin{equation}
\boldsymbol{k} = \begin{bmatrix}
6.59 \\
3.22 \\
1.37
\end{bmatrix} \times 10^{-9} \, ,
\end{equation}
where we fix the ratio of scale lengths $h/l = 0.02$.
%vertical scale and radial scale lengths $h = 0.01$, and $l = 0.5$,  respectively. 
Furthermore, for model 1, we present the magnetic field structure for vertical-to-radial scale length ratio $h/l \in [0.01, 0.1, 0.75]$. %Defining $(h/l_1, h/l_2, h/l_3) = (0.01, 0.1, 0.75)$, we can compute the respective vectors of coefficients:
For these three values of $h/l$, we compute the respective vector of coefficients to be
\begin{equation}
\begin{aligned}
&\boldsymbol{k}^{(1)} = \begin{bmatrix}
3.29 \\
1.61 \\
0.68
\end{bmatrix} \times 10^{-9}, \quad
\boldsymbol{k}^{(2)} = \begin{bmatrix}
3.32 \\
1.66 \\
0.72
\end{bmatrix} \times 10^{-8}, \\
&\quad \quad \quad \quad \quad \quad \boldsymbol{k}^{(3)} = \begin{bmatrix}
4.14 \\
6.81 \\
19.52
\end{bmatrix} \times 10^{-7} \, .
\end{aligned}
\end{equation}
%corresponding to $(h/l_1, h/l_2, h/l_3) = (0.01, 0.1, 0.75)$.
Here, $\boldsymbol{k}^{(1)}$ is computed using $h/l = 0.01$, $\boldsymbol{k}^{(2)}$ is computed using $h/l = 0.1$, and $\boldsymbol{k}^{(3)}$ is computed using $h/l = 0.75$. 
Figure \ref{model_1b} shows the magnetic field line morphology in all three cases for a two-dimensional slice through the center of a model. Figure \ref{model_1_3d} shows a three-dimensional rendering of the axisymmetric models.
Moving onto model 2, we take a Bessel function for the radial current density distribution and vary the vertical scale length $h$. We note that for the Bessel distribution, the core radius $R$ serves as the scale length. We present values of $h/R \in [0.01,0.025,0.1]$.  The vector of coefficients is calculated to be
\begin{equation}
\begin{aligned}
&\boldsymbol{k}^{(1)} = \begin{bmatrix}
11.61 \\
2.63 \\
1.18
\end{bmatrix} \times 10^{-9}, \quad
\boldsymbol{k}^{(2)} = \begin{bmatrix}
29.09 \\
6.63 \\
2.99
\end{bmatrix} \times 10^{-9}, \\
&\quad \quad \quad \quad \quad \quad \boldsymbol{k}^{(3)} = \begin{bmatrix}
12.04 \\
2.98 \\
1.53
\end{bmatrix} \times 10^{-8},
\end{aligned}
\end{equation}
where the meaning of each $\boldsymbol{k}^{(i)}$ is as used before.
Figure \ref{model_2} shows that for increasing  ratio $h/R$, there is stronger curvature in the field lines and eventually there are some closed loops for the case $h/R = 0.1$. Figure \ref{model_2_3d} shows the corresponding three-dimensional renderings of the field lines.
%corresponding to $(h/R_1, h/R_2, h/R_3) = (0.01, 0.025, 0.1)$. 
%Finally, we represent the radial current  profile by the Power-Law function.
Finally, we present model 3, which has the power-law radial current density profile.
%By surveying the ratio of $h/l \in [0.01, 0.1, 0.75]$ as done for model (1), we demonstrate in Figure \ref{model_3}, how for increasing $h/l$, the field morphology approaches a more toroidal like structure. 
 We evaluate the model for values of the ratio $h/l \in [0.01, 0.1, 0.75]$, as done for model 1. 
%Defining again $(h/l_1, h/l_2, h/l_3) = (0.01, 0.1, 0.75)$, we compute the respective vectors of coefficients
For this model, the respective vector of coefficients is calculated to be
\begin{equation}
\begin{aligned}
&\boldsymbol{k}^{(1)} = \begin{bmatrix}
48.58 \\
8.92 \\
9.31
\end{bmatrix} \times 10^{-10}, \quad
\boldsymbol{k}^{(2)} = \begin{bmatrix}
49.02 \\
9.19 \\
9.92
\end{bmatrix} \times 10^{-9}, \\
&\quad \quad \quad \quad \quad \quad \boldsymbol{k}^{(3)} = \begin{bmatrix}
6.10 \\
3.77 \\
26.54
\end{bmatrix} \times 10^{-7}.
\end{aligned}
\end{equation}
%corresponding to $(h/l_1, h/l_2, h/l_3) = (0.01, 0.1, 0.75)$.
In Figure \ref{model_3}, we see that the field line morphology again approaches a more pinched structure with increasing $h/l$. Figure \ref{model_3_3d} shows the three-dimensional rendering of the field lines for these models.
\section{Discussion}
\label{discussion}
\begin{figure*}
\centering
\hspace{-0.6cm}
  \includegraphics[width=1\textwidth]{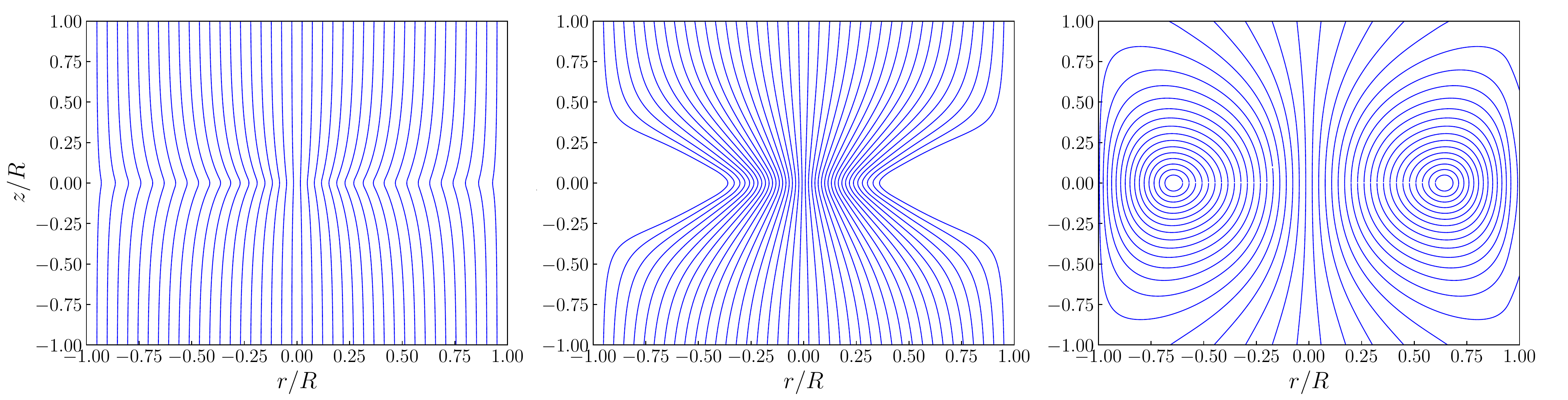}
  \vspace{-0.3cm}
  \caption{Magnetic field morphology for model 3 (power-law radial and Gaussian vertical distributions) demonstrated for the scale length ratio $h/l \in [0.01, 0.1, 0.75]$ (from left to right) having respective $B_c/B_0 \in [3.07, 17.21, 61.93]$.}
 \label{model_3}
  \vspace{0.5cm}
   \includegraphics[width=1\textwidth]{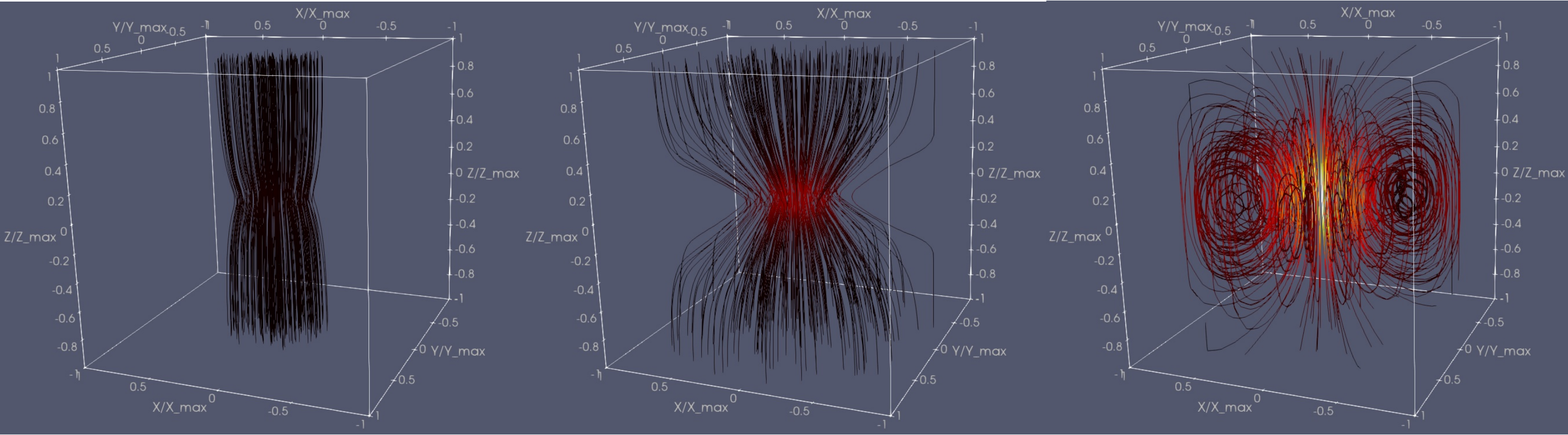}
  \caption{Three-dimensional renderings of the magnetic field morphology in Figure \ref{model_3}. The color gradient is representative of the field strength ranging from the black being weakest to yellow being strongest.}
  \label{model_3_3d}
\end{figure*}

Upon visual inspection of the magnetic field structures shown in Figures \ref{model_1a} to \ref{model_3_3d}, %\ref{model_1b}, \ref{model_2}, \ref{model_3}, 
we see that the methodology presented in this paper can be used to model a wide array of magnetic field morphologies. 
%Additionally, in Figures \ref{model_1_3d}, \ref{model_2_3d}, and \ref{model_3_3d} we include  the three dimensional rendering of the field lines for each respective model. 
%The structures of these figures are assumed axisymmetric about the azimuthal angle.
%({\bf These figures are obtained assuming azimuthal symmetry})  
%rotationally axisymmetric about $z-$axis
We note, however, that the selection of current density profiles is not limited to the ones introduced here. 
%but rather any centered bell shaped distributions can be used to model the current density distribution. 
By varying the model parameters (e.g., $h$ and $l$), 
%we can manipulate the concentration of the current along their respective dimensions.
the concentration of the current density can be adjusted along the respective directions.
%In doing such, we see that when we increase the characteristic scale length along a given direction, the current becomes more distributed over its respective domain.
Simulations of magnetized fragmentation \citep[e.g.,][]{kudohbasu2007,kudoh2011formation} suggest that a smaller $h/l$ is usually preferred because a relatively slow gravitational contraction gives gas enough time to settle along the direction of the field lines. However, if dynamical contraction is fast enough to prevent gas from settling along the vertical direction, we may observe systems exhibiting higher $h/l$ ratios. A rapid core formation induced by a large-scale nonlinear flow may lead to this kind of situation \cite[][see \S\ 4]{kudohbasu2008}, in which the vertical equilibrium along the mean magnetic field cannot be established in the time available.
The solutions with small $B_0$, i.e., weak background field, also exhibit loop-like magnetic fields. Physically, this means that there has been enough contraction that the induced current dominates the effect of external currents. 
The cases with larger (order unity) $h/l$ and low $B_0$ may both represent the limit of fast contraction of a core from a distance that is larger than that which may be associated with the gas density of an observed dense core. \citet{binobasu2021} found that a solution with large field line curvature provided a good fit to the polarization segments measured in FeSt 1--457, implying a contraction from a larger distance than normally associated with the gas density structure of the core. If the length scales associated with the magnetic field and the gas density are different, it may be taken to imply the breakdown of flux-freezing (i.e., ambipolar diffusion) during the core formation process. The structure of the magnetic field modeled in  \citet{binobasu2021} is similar to those demonstrated in the third panel of Figure \ref{model_2} for model 2 and the second panel of Figure \ref{model_3} for model 3.
%This closed loop morphology has been used to model the magnetic state of the prestellar core FeSt 1--457 \citep{binobasu2021}. The structure of the magnetic field modeled in \cite{binobasu2021} is similar to those demonstrated in the third panel of Figure \ref{model_2} for model 2 and the second panel of Figure \ref{model_3} for model 3. Physically, the strong curvature in the magnetic field suggests that the core had contracted from a significantly larger radius than that observed. This is believed to have been caused by a rapid nonlinear flow \citep{binobasu2021}.

In the protostellar phase, magnetic loops can also arise because of redistribution of magnetic flux by reconnecting field lines \citep[e.g.,][]{Surianoetal2019MNRAS}. This might happen near the protostar because of high magnetic diffusion (Ohmic diffusivity) leading to reconnection of oppositely directed field lines. 
As the field lines are dragged inward, oppositely directed field lines can converge to the equatorial plane and then reconnect, leading to closed field line configurations. In this case, the 
%because of the bending and form a loop, and the loop gradually sinks. This is how we may think that the magnetic flux can be dissipated. For this situation, the 
current density also becomes concentrated near the center. 

%SB: commented out below as it was redundant
%These scenarios are expected to exhibit field line structures similar to the right hand plot in Figure \ref{model_1b}, assuming the background field is relatively weak enough. Additionally, we can see that the normalized central magnetic field value $B_c/B_0$ plays a large role in the field morphology. Particularly, we take this value normalized to either the interstellar or intercore medium in most circumstances.
\section{Conclusion}
\label{conclusion}
In this paper, we have provided an extension to the mathematical model first introduced by \cite{ewertowski2013mathematical} and deployed to model observed data of a prestellar core \citep{binobasu2021}. Rather than treating the coefficients $k_m$ as parameters used for fitting, we directly computed them by proposing a current density profile along the radial direction. The adopted functional forms (Gaussian, Bessel, and power law) are plausible examples of centrally-concentrated profiles that emerge from analytical and numerical models of magnetized contraction. We implemented a variety of such functions for the distribution of current density over the $r$-direction, while maintaining a Gaussian function over the $z$-direction. Our findings demonstrated a variety of hourglass morphologies. We studied the model's response to the parameters by varying the respective scale lengths and the background field. Interestingly, a very weak background field $B_0$ is not the only cause of a very pinched hourglass structure. We found that the morphology of the magnetic field is also dictated by the ratio $h/l$ between the characteristic scale lengths along the $z$- and $r$- directions. Our calculations indicated that increasing the ratio $h/l$ yields an increasingly more pinched hourglass shape. 
%Physically, increasing the characteristic scale length along any given direction is equivalent to redistributing more current around the central region. This directly reduces the amount of current closer to the boundaries which weakens the magnetic field in these areas. This result provides an interesting and alternative methodology to the analysis of hourglass magnetic field. 
We conclude that we cannot directly assume that strong curvature in the field lines is due to a weak $B_0$, but rather that it can also be caused by how the current density is distributed throughout the core. 
%Additionally, if we are assuming the profile for the current along the radial direction, we indirectly reduce the number of parameters to fit from $m + 3$ down to a maximum of 4 for $h$, $l$, $R$ and $B_0$ (this number can be further reduced to 3 if we have an estimate of the core radius $R$). 

When we adopt a specific form for the radial distribution $f(r)$ of current density, the number of parameters is a maximum of four ($h$, $l$, $R$ and $B_0$). This can be an advantage over the original method of \cite{ewertowski2013mathematical}, in which the number of fitting parameters increases with the number of terms utilized in the series expansion solution of equations (\ref{eq:A_series2}), (\ref{eq:Br2}), and (\ref{eq:Bz2}).
Our modified technique requires direct computation of the integral in equation (\ref{eq:4}), but the optimization problem is potentially made easier by a reduction of the number of fitting parameters.
%We have surveyed three potential profiles to model the current along the radial direction: the Gaussian, Bessel, and power-law functions. Each profile gave rise to a series of hourglass figures, and they all required a form of numerical integration in order to compute the coefficients in equation (\ref{eq:4}). 
The machinery developed in this paper can be used to efficiently model the hourglass profiles inferred from polarimetry observations of star-forming regions. Future work can investigate other functional forms of $f(r)$, including ones that may lead to patterns that are not in an hourglass form. Additionally, it will be useful to seek current density distributions for which the coefficients $k_m$ can be analytically evaluated.\\ \\
%We thank the referee for constructive comments. 
We thank the Department of Applied Mathematics at U. W. O. for providing a hospitable environment that allowed the authors to form a fruitful collaboration. S. B. is supported by a Discovery Grant from NSERC.

%\section*{References}

\bibliography{manuscript}

\begin{thebibliography}{27}
\expandafter\ifx\csname natexlab\endcsname\relax\def\natexlab#1{#1}\fi
\providecommand{\url}[1]{\texttt{#1}}
\providecommand{\href}[2]{#2}
\providecommand{\path}[1]{#1}
\providecommand{\DOIprefix}{doi:}
\providecommand{\ArXivprefix}{arXiv:}
\providecommand{\URLprefix}{URL: }
\providecommand{\Pubmedprefix}{pmid:}
\providecommand{\doi}[1]{\href{http://dx.doi.org/#1}{\path{#1}}}
\providecommand{\Pubmed}[1]{\href{pmid:#1}{\path{#1}}}
\providecommand{\bibinfo}[2]{#2}
\ifx\xfnm\relax \def\xfnm[#1]{\unskip,\space#1}\fi
%Type = Article
\bibitem[{{Basu} et~al.(2009){Basu}, {Ciolek} and {Wurster}}]{bas09}
\bibinfo{author}{{Basu}, S.}, \bibinfo{author}{{Ciolek}, G.E.},
  \bibinfo{author}{{Wurster}, J.}, \bibinfo{year}{2009}.
\newblock \bibinfo{title}{Nonlinear evolution of gravitational fragmentation
  regulated by magnetic fields and ambipolar diffusion}.
\newblock \bibinfo{journal}{New Astronomy} \bibinfo{volume}{14},
  \bibinfo{pages}{221--237}.
\newblock \DOIprefix\doi{10.1016/j.newast.2008.07.006},
  \href{http://arxiv.org/abs/0806.2482}{\tt arXiv:0806.2482}.
%Type = Article
\bibitem[{{Bino} and {Basu}(2021)}]{binobasu2021}
\bibinfo{author}{{Bino}, G.}, \bibinfo{author}{{Basu}, S.},
  \bibinfo{year}{2021}.
\newblock \bibinfo{title}{Fitting an analytic magnetic field to a prestellar
  core}.
\newblock \bibinfo{journal}{Astrophysical Journal} \bibinfo{volume}{911},
  \bibinfo{pages}{15}.
\newblock \DOIprefix\doi{10.3847/1538-4357/abe6a4},
  \href{http://arxiv.org/abs/2103.03324}{\tt arXiv:2103.03324}.
%Type = Article
\bibitem[{{Chandrasekhar} and {Fermi}(1953)}]{CF1953}
\bibinfo{author}{{Chandrasekhar}, S.}, \bibinfo{author}{{Fermi}, E.},
  \bibinfo{year}{1953}.
\newblock \bibinfo{title}{Problems of gravitational stability in the presence
  of a magnetic field.}
\newblock \bibinfo{journal}{Astrophysical Journal} \bibinfo{volume}{118},
  \bibinfo{pages}{116}.
\newblock \DOIprefix\doi{10.1086/145732}.
%Type = Article
\bibitem[{{Davis}(1951)}]{davis1951}
\bibinfo{author}{{Davis}, L.}, \bibinfo{year}{1951}.
\newblock \bibinfo{title}{The strength of interstellar magnetic fields}.
\newblock \bibinfo{journal}{Physical Review} \bibinfo{volume}{81},
  \bibinfo{pages}{890--891}.
\newblock \DOIprefix\doi{10.1103/PhysRev.81.890.2}.
%Type = Article
\bibitem[{{Ewertowski} and {Basu}(2013)}]{ewertowski2013mathematical}
\bibinfo{author}{{Ewertowski}, B.}, \bibinfo{author}{{Basu}, S.},
  \bibinfo{year}{2013}.
\newblock \bibinfo{title}{{A Mathematical Model for an Hourglass Magnetic
  Field}}.
\newblock \bibinfo{journal}{\apj} \bibinfo{volume}{767}, \bibinfo{pages}{33}.
\newblock \DOIprefix\doi{10.1088/0004-637X/767/1/33},
  \href{http://arxiv.org/abs/1302.6913}{\tt arXiv:1302.6913}.
%Type = Article
\bibitem[{{Frau} et~al.(2011){Frau}, {Galli} and {Girart}}]{frau2011}
\bibinfo{author}{{Frau}, P.}, \bibinfo{author}{{Galli}, D.},
  \bibinfo{author}{{Girart}, J.M.}, \bibinfo{year}{2011}.
\newblock \bibinfo{title}{Comparing star formation models with interferometric
  observations of the protostar ngc 1333 iras 4a. i. magnetohydrodynamic
  collapse models}.
\newblock \bibinfo{journal}{Astronomy \& Astrophysics} \bibinfo{volume}{535},
  \bibinfo{pages}{A44}.
\newblock \DOIprefix\doi{10.1051/0004-6361/201117813},
  \href{http://arxiv.org/abs/1109.6251}{\tt arXiv:1109.6251}.
%Type = Article
\bibitem[{{Girart} et~al.(2009){Girart}, {Beltr{\'a}n}, {Zhang}, {Rao} and
  {Estalella}}]{Giratetal2009}
\bibinfo{author}{{Girart}, J.M.}, \bibinfo{author}{{Beltr{\'a}n}, M.T.},
  \bibinfo{author}{{Zhang}, Q.}, \bibinfo{author}{{Rao}, R.},
  \bibinfo{author}{{Estalella}, R.}, \bibinfo{year}{2009}.
\newblock \bibinfo{title}{Magnetic fields in the formation of massive stars}.
\newblock \bibinfo{journal}{Science} \bibinfo{volume}{324},
  \bibinfo{pages}{1408}.
\newblock \DOIprefix\doi{10.1126/science.1171807}.
%Type = Article
\bibitem[{{Girart} et~al.(2006){Girart}, {Rao} and {Marrone}}]{Giratetal2006}
\bibinfo{author}{{Girart}, J.M.}, \bibinfo{author}{{Rao}, R.},
  \bibinfo{author}{{Marrone}, D.P.}, \bibinfo{year}{2006}.
\newblock \bibinfo{title}{Magnetic fields in the formation of sun-like stars}.
\newblock \bibinfo{journal}{Science} \bibinfo{volume}{313},
  \bibinfo{pages}{812--814}.
\newblock \DOIprefix\doi{10.1126/science.1129093},
  \href{http://arxiv.org/abs/astro-ph/0609177}{\tt arXiv:astro-ph/0609177}.
%Type = Article
\bibitem[{{Kandori} et~al.(2018){Kandori}, {Tamura}, {Nagata}, {Tomisaka},
  {Kusakabe}, {Nakajima}, {Kwon}, {Nagayama} and {Tatematsu}}]{kan3}
\bibinfo{author}{{Kandori}, R.}, \bibinfo{author}{{Tamura}, M.},
  \bibinfo{author}{{Nagata}, T.}, \bibinfo{author}{{Tomisaka}, K.},
  \bibinfo{author}{{Kusakabe}, N.}, \bibinfo{author}{{Nakajima}, Y.},
  \bibinfo{author}{{Kwon}, J.}, \bibinfo{author}{{Nagayama}, T.},
  \bibinfo{author}{{Tatematsu}, K.}, \bibinfo{year}{2018}.
\newblock \bibinfo{title}{Distortion of magnetic fields in a starless core.
  iii. polarization-extinction relationship in fest 1-457}.
\newblock \bibinfo{journal}{Astrophysical Journal} \bibinfo{volume}{857},
  \bibinfo{pages}{100}.
\newblock \DOIprefix\doi{10.3847/1538-4357/aab962},
  \href{http://arxiv.org/abs/1803.08196}{\tt arXiv:1803.08196}.
%Type = Article
\bibitem[{{Kandori} et~al.(2017){Kandori}, {Tamura}, {Tomisaka}, {Nakajima},
  {Kusakabe}, {Kwon}, {Nagayama}, {Nagata} and {Tatematsu}}]{Kandori_2017}
\bibinfo{author}{{Kandori}, R.}, \bibinfo{author}{{Tamura}, M.},
  \bibinfo{author}{{Tomisaka}, K.}, \bibinfo{author}{{Nakajima}, Y.},
  \bibinfo{author}{{Kusakabe}, N.}, \bibinfo{author}{{Kwon}, J.},
  \bibinfo{author}{{Nagayama}, T.}, \bibinfo{author}{{Nagata}, T.},
  \bibinfo{author}{{Tatematsu}, K.}, \bibinfo{year}{2017}.
\newblock \bibinfo{title}{Distortion of magnetic fields in a starless core ii:
  3d magnetic field structure of fest 1-457}.
\newblock \bibinfo{journal}{Astrophysical Journal} \bibinfo{volume}{848},
  \bibinfo{pages}{110}.
\newblock \DOIprefix\doi{10.3847/1538-4357/aa8d18},
  \href{http://arxiv.org/abs/1709.04544}{\tt arXiv:1709.04544}.
%Type = Article
\bibitem[{{Kataoka} et~al.(2012){Kataoka}, {Machida} and
  {Tomisaka}}]{kataoka2012}
\bibinfo{author}{{Kataoka}, A.}, \bibinfo{author}{{Machida}, M.N.},
  \bibinfo{author}{{Tomisaka}, K.}, \bibinfo{year}{2012}.
\newblock \bibinfo{title}{Exploring magnetic field structure in star-forming
  cores with polarization of thermal dust emission}.
\newblock \bibinfo{journal}{Astrophysical Journal} \bibinfo{volume}{761},
  \bibinfo{pages}{40}.
\newblock \DOIprefix\doi{10.1088/0004-637X/761/1/40},
  \href{http://arxiv.org/abs/1210.7637}{\tt arXiv:1210.7637}.
%Type = Article
\bibitem[{{Kudoh} and {Basu}(2008)}]{kudohbasu2008}
\bibinfo{author}{{Kudoh}, T.}, \bibinfo{author}{{Basu}, S.},
  \bibinfo{year}{2008}.
\newblock \bibinfo{title}{Three-dimensional simulation of magnetized cloud
  fragmentation induced by nonlinear flows and ambipolar diffusion}.
\newblock \bibinfo{journal}{\apjl} \bibinfo{volume}{679}, \bibinfo{pages}{L97}.
\newblock \DOIprefix\doi{10.1086/589618},
  \href{http://arxiv.org/abs/0804.4303}{\tt arXiv:0804.4303}.
%Type = Article
\bibitem[{{Kudoh} and {Basu}(2011)}]{kudoh2011formation}
\bibinfo{author}{{Kudoh}, T.}, \bibinfo{author}{{Basu}, S.},
  \bibinfo{year}{2011}.
\newblock \bibinfo{title}{Formation of collapsing cores in subcritical magnetic
  clouds: Three-dimensional magnetohydrodynamic simulations with ambipolar
  diffusion}.
\newblock \bibinfo{journal}{Astrophysical Journal} \bibinfo{volume}{728},
  \bibinfo{pages}{123}.
\newblock \DOIprefix\doi{10.1088/0004-637X/728/2/123},
  \href{http://arxiv.org/abs/1012.5707}{\tt arXiv:1012.5707}.
%Type = Article
\bibitem[{{Kudoh} et~al.(2007){Kudoh}, {Basu}, {Ogata} and
  {Yabe}}]{kudohbasu2007}
\bibinfo{author}{{Kudoh}, T.}, \bibinfo{author}{{Basu}, S.},
  \bibinfo{author}{{Ogata}, Y.}, \bibinfo{author}{{Yabe}, T.},
  \bibinfo{year}{2007}.
\newblock \bibinfo{title}{Three-dimensional simulations of molecular cloud
  fragmentation regulated by magnetic fields and ambipolar diffusion}.
\newblock \bibinfo{journal}{Monthly Notices of the Royal Astronomical Society}
  \bibinfo{volume}{380}, \bibinfo{pages}{499--505}.
\newblock \DOIprefix\doi{10.1111/j.1365-2966.2007.12119.x},
  \href{http://arxiv.org/abs/0706.2696}{\tt arXiv:0706.2696}.
%Type = Article
\bibitem[{{Kwon} et~al.(2019){Kwon}, {Stephens}, {Tobin}, {Looney}, {Li}, {van
  der Tak} and {Crutcher}}]{kwon2019}
\bibinfo{author}{{Kwon}, W.}, \bibinfo{author}{{Stephens}, I.W.},
  \bibinfo{author}{{Tobin}, J.J.}, \bibinfo{author}{{Looney}, L.W.},
  \bibinfo{author}{{Li}, Z.Y.}, \bibinfo{author}{{van der Tak}, F.F.S.},
  \bibinfo{author}{{Crutcher}, R.M.}, \bibinfo{year}{2019}.
\newblock \bibinfo{title}{Highly ordered and pinched magnetic fields in the
  class 0 protobinary system l1448 irs 2}.
\newblock \bibinfo{journal}{Astrophysical Journal} \bibinfo{volume}{879},
  \bibinfo{pages}{25}.
\newblock \DOIprefix\doi{10.3847/1538-4357/ab24c8},
  \href{http://arxiv.org/abs/1805.07348}{\tt arXiv:1805.07348}.
%Type = Article
\bibitem[{Mestel and Spitzer(1956)}]{mestel1956}
\bibinfo{author}{Mestel, L.}, \bibinfo{author}{Spitzer, L., J.},
  \bibinfo{year}{1956}.
\newblock \bibinfo{title}{Star formation in magnetic dust clouds}.
\newblock \bibinfo{journal}{Monthly Notices of the Royal Astronomical Society}
  \bibinfo{volume}{116}, \bibinfo{pages}{503}.
\newblock \DOIprefix\doi{10.1093/mnras/116.5.503}.
%Type = Inproceedings
\bibitem[{{Mouschovias}(1978)}]{mouschovias78}
\bibinfo{author}{{Mouschovias}, T.C.}, \bibinfo{year}{1978}.
\newblock \bibinfo{title}{Formation of stars and planetary systems in magnetic
  interstellar clouds}, in: \bibinfo{editor}{{Gehrels}, T.},
  \bibinfo{editor}{{Matthews}, M.S.} (Eds.), \bibinfo{booktitle}{IAU Colloq.
  52: Protostars and Planets}, \bibinfo{publisher}{University of Arizona
  Press}. p. \bibinfo{pages}{209}.
%Type = Inproceedings
\bibitem[{{Mouschovias} and {Ciolek}(1999)}]{mou1999}
\bibinfo{author}{{Mouschovias}, T.C.}, \bibinfo{author}{{Ciolek}, G.E.},
  \bibinfo{year}{1999}.
\newblock \bibinfo{title}{Magnetic fields and star formation: A theory reaching
  adulthood}, in: \bibinfo{editor}{{Lada}, C.J.}, \bibinfo{editor}{{Kylafis},
  N.D.} (Eds.), \bibinfo{booktitle}{NATO Advanced Science Institutes (ASI)
  Series C}, \bibinfo{publisher}{Kluwer Academic Publishers}. p.
  \bibinfo{pages}{305}.
%Type = Article
\bibitem[{{Pattle} et~al.(2017){Pattle}, {Ward-Thompson}, {Berry}, {Hatchell},
  {Chen}, {Pon}, {Koch}, {Kwon}, {Kim}, {Bastien}, {Cho}, {Coud{\'e}}, {Di
  Francesco}, {Fuller}, {Furuya}, {Graves}, {Johnstone}, {Kirk}, {Kwon}, {Lee},
  {Matthews}, {Mottram}, {Parsons}, {Sadavoy}, {Shinnaga}, {Soam}, {Hasegawa},
  {Lai}, {Qiu} and {Friberg}}]{Pattle2017}
\bibinfo{author}{{Pattle}, K.}, \bibinfo{author}{{Ward-Thompson}, D.},
  \bibinfo{author}{{Berry}, D.}, \bibinfo{author}{{Hatchell}, J.},
  \bibinfo{author}{{Chen}, H.R.}, \bibinfo{author}{{Pon}, A.},
  \bibinfo{author}{{Koch}, P.M.}, \bibinfo{author}{{Kwon}, W.},
  \bibinfo{author}{{Kim}, J.}, \bibinfo{author}{{Bastien}, P.},
  \bibinfo{author}{{Cho}, J.}, \bibinfo{author}{{Coud{\'e}}, S.},
  \bibinfo{author}{{Di Francesco}, J.}, \bibinfo{author}{{Fuller}, G.},
  \bibinfo{author}{{Furuya}, R.S.}, \bibinfo{author}{{Graves}, S.F.},
  \bibinfo{author}{{Johnstone}, D.}, \bibinfo{author}{{Kirk}, J.},
  \bibinfo{author}{{Kwon}, J.}, \bibinfo{author}{{Lee}, C.W.},
  \bibinfo{author}{{Matthews}, B.C.}, \bibinfo{author}{{Mottram}, J.C.},
  \bibinfo{author}{{Parsons}, H.}, \bibinfo{author}{{Sadavoy}, S.},
  \bibinfo{author}{{Shinnaga}, H.}, \bibinfo{author}{{Soam}, A.},
  \bibinfo{author}{{Hasegawa}, T.}, \bibinfo{author}{{Lai}, S.P.},
  \bibinfo{author}{{Qiu}, K.}, \bibinfo{author}{{Friberg}, P.},
  \bibinfo{year}{2017}.
\newblock \bibinfo{title}{The jcmt bistro survey: The magnetic field strength
  in the orion a filament}.
\newblock \bibinfo{journal}{Astrophysical Journal} \bibinfo{volume}{846},
  \bibinfo{pages}{122}.
\newblock \DOIprefix\doi{10.3847/1538-4357/aa80e5},
  \href{http://arxiv.org/abs/1707.05269}{\tt arXiv:1707.05269}.
%Type = Article
\bibitem[{{Planck Collaboration} et~al.(2015){Planck Collaboration}, {Ade},
  {Aghanim}, {Alina}, {Aniano}, {Armitage-Caplan}, {Arnaud}, {Ashdown},
  {Atrio-Barandela}, {Aumont}, {Baccigalupi}, {Banday}, {Barreiro}, {Battaner},
  {Beichman}, {Benabed}, {Benoit-L{\'e}vy}, {Bernard}, {Bersanelli},
  {Bielewicz}, {Bock}, {Bond}, {Borrill}, {Bouchet}, {Boulanger}, {Burigana},
  {Cardoso}, {Catalano}, {Chamballu}, {Chary}, {Chiang}, {Christensen},
  {Colombi}, {Colombo}, {Combet}, {Couchot}, {Coulais}, {Crill}, {Curto},
  {Cuttaia}, {Danese}, {Davies}, {Davis}, {de Bernardis}, {de Rosa}, {de
  Zotti}, {Delabrouille}, {D{\'e}sert}, {Dickinson}, {Diego}, {Donzelli},
  {Dor{\'e}}, {Douspis}, {Dunkley}, {Dupac}, {Efstathiou}, {En{\ss}lin},
  {Eriksen}, {Falgarone}, {Fanciullo}, {Finelli}, {Forni}, {Frailis},
  {Fraisse}, {Franceschi}, {Galeotta}, {Ganga}, {Ghosh}, {Giard},
  {Giraud-H{\'e}raud}, {Gonz{\'a}lez-Nuevo}, {G{\'o}rski}, {Gregorio},
  {Gruppuso}, {Guillet}, {Hansen}, {Harrison}, {Helou},
  {Hern{\'a}ndez-Monteagudo}, {Hildebrand t}, {Hivon}, {Hobson}, {Holmes},
  {Hornstrup}, {Huffenberger}, {Jaffe}, {Jaffe}, {Jones}, {Juvela},
  {Keih{\"a}nen}, {Keskitalo}, {Kisner}, {Kneissl}, {Knoche}, {Kunz},
  {Kurki-Suonio}, {Lagache}, {L{\"a}hteenm{\"a}ki}, {Lamarre}, {Lasenby},
  {Lawrence}, {Leonardi}, {Levrier}, {Liguori}, {Lilje}, {Linden-V{\o}rnle},
  {L{\'o}pez-Caniego}, {Lubin}, {Mac{\'\i}as-P{\'e}rez}, {Maffei},
  {Magalh{\~a}es}, {Maino}, {Mandolesi}, {Maris}, {Marshall}, {Martin},
  {Mart{\'\i}nez-Gonz{\'a}lez}, {Masi}, {Matarrese}, {Mazzotta}, {Melchiorri},
  {Mendes}, {Mennella}, {Migliaccio}, {Miville-Desch{\^e}nes}, {Moneti},
  {Montier}, {Morgante}, {Mortlock}, {Munshi}, {Murphy}, {Naselsky}, {Nati},
  {Natoli}, {Netterfield}, {Noviello}, {Novikov}, {Novikov}, {Oxborrow},
  {Pagano}, {Pajot}, {Paladini}, {Paoletti}, {Pasian}, {Perdereau}, {Perotto},
  {Perrotta}, {Piacentini}, {Piat}, {Pietrobon}, {Plaszczynski}, {Poidevin},
  {Pointecouteau}, {Polenta}, {Popa}, {Pratt}, {Prunet}, {Puget}, {Rachen},
  {Reach}, {Rebolo}, {Reinecke}, {Remazeilles}, {Renault}, {Ricciardi},
  {Riller}, {Ristorcelli}, {Rocha}, {Rosset}, {Roudier}, {Rusholme}, {Sandri},
  {Savini}, {Scott}, {Spencer}, {Stolyarov}, {Stompor}, {Sudiwala}, {Sutton},
  {Suur-Uski}, {Sygnet}, {Tauber}, {Terenzi}, {Toffolatti}, {Tomasi},
  {Tristram}, {Tucci}, {Umana}, {Valenziano}, {Valiviita}, {Van Tent},
  {Vielva}, {Villa}, {Wade}, {Wandelt} and {Zonca}}]{planck2015}
\bibinfo{author}{{Planck Collaboration}}, \bibinfo{author}{{Ade}, P.A.R.},
  \bibinfo{author}{{Aghanim}, N.}, \bibinfo{author}{{Alina}, D.},
  \bibinfo{author}{{Aniano}, G.}, \bibinfo{author}{{Armitage-Caplan}, C.},
  \bibinfo{author}{{Arnaud}, M.}, \bibinfo{author}{{Ashdown}, M.},
  \bibinfo{author}{{Atrio-Barandela}, F.}, \bibinfo{author}{{Aumont}, J.},
  \bibinfo{author}{{Baccigalupi}, C.}, \bibinfo{author}{{Banday}, A.J.},
  \bibinfo{author}{{Barreiro}, R.B.}, \bibinfo{author}{{Battaner}, E.},
  \bibinfo{author}{{Beichman}, C.}, \bibinfo{author}{{Benabed}, K.},
  \bibinfo{author}{{Benoit-L{\'e}vy}, A.}, \bibinfo{author}{{Bernard}, J.P.},
  \bibinfo{author}{{Bersanelli}, M.}, \bibinfo{author}{{Bielewicz}, P.},
  \bibinfo{author}{{Bock}, J.J.}, \bibinfo{author}{{Bond}, J.R.},
  \bibinfo{author}{{Borrill}, J.}, \bibinfo{author}{{Bouchet}, F.R.},
  \bibinfo{author}{{Boulanger}, F.}, \bibinfo{author}{{Burigana}, C.},
  \bibinfo{author}{{Cardoso}, J.F.}, \bibinfo{author}{{Catalano}, A.},
  \bibinfo{author}{{Chamballu}, A.}, \bibinfo{author}{{Chary}, R.R.},
  \bibinfo{author}{{Chiang}, H.C.}, \bibinfo{author}{{Christensen}, P.R.},
  \bibinfo{author}{{Colombi}, S.}, \bibinfo{author}{{Colombo}, L.P.L.},
  \bibinfo{author}{{Combet}, C.}, \bibinfo{author}{{Couchot}, F.},
  \bibinfo{author}{{Coulais}, A.}, \bibinfo{author}{{Crill}, B.P.},
  \bibinfo{author}{{Curto}, A.}, \bibinfo{author}{{Cuttaia}, F.},
  \bibinfo{author}{{Danese}, L.}, \bibinfo{author}{{Davies}, R.D.},
  \bibinfo{author}{{Davis}, R.J.}, \bibinfo{author}{{de Bernardis}, P.},
  \bibinfo{author}{{de Rosa}, A.}, \bibinfo{author}{{de Zotti}, G.},
  \bibinfo{author}{{Delabrouille}, J.}, \bibinfo{author}{{D{\'e}sert}, F.X.},
  \bibinfo{author}{{Dickinson}, C.}, \bibinfo{author}{{Diego}, J.M.},
  \bibinfo{author}{{Donzelli}, S.}, \bibinfo{author}{{Dor{\'e}}, O.},
  \bibinfo{author}{{Douspis}, M.}, \bibinfo{author}{{Dunkley}, J.},
  \bibinfo{author}{{Dupac}, X.}, \bibinfo{author}{{Efstathiou}, G.},
  \bibinfo{author}{{En{\ss}lin}, T.A.}, \bibinfo{author}{{Eriksen}, H.K.},
  \bibinfo{author}{{Falgarone}, E.}, \bibinfo{author}{{Fanciullo}, L.},
  \bibinfo{author}{{Finelli}, F.}, \bibinfo{author}{{Forni}, O.},
  \bibinfo{author}{{Frailis}, M.}, \bibinfo{author}{{Fraisse}, A.A.},
  \bibinfo{author}{{Franceschi}, E.}, \bibinfo{author}{{Galeotta}, S.},
  \bibinfo{author}{{Ganga}, K.}, \bibinfo{author}{{Ghosh}, T.},
  \bibinfo{author}{{Giard}, M.}, \bibinfo{author}{{Giraud-H{\'e}raud}, Y.},
  \bibinfo{author}{{Gonz{\'a}lez-Nuevo}, J.}, \bibinfo{author}{{G{\'o}rski},
  K.M.}, \bibinfo{author}{{Gregorio}, A.}, \bibinfo{author}{{Gruppuso}, A.},
  \bibinfo{author}{{Guillet}, V.}, \bibinfo{author}{{Hansen}, F.K.},
  \bibinfo{author}{{Harrison}, D.L.}, \bibinfo{author}{{Helou}, G.},
  \bibinfo{author}{{Hern{\'a}ndez-Monteagudo}, C.},
  \bibinfo{author}{{Hildebrand t}, S.R.}, \bibinfo{author}{{Hivon}, E.},
  \bibinfo{author}{{Hobson}, M.}, \bibinfo{author}{{Holmes}, W.A.},
  \bibinfo{author}{{Hornstrup}, A.}, \bibinfo{author}{{Huffenberger}, K.M.},
  \bibinfo{author}{{Jaffe}, A.H.}, \bibinfo{author}{{Jaffe}, T.R.},
  \bibinfo{author}{{Jones}, W.C.}, \bibinfo{author}{{Juvela}, M.},
  \bibinfo{author}{{Keih{\"a}nen}, E.}, \bibinfo{author}{{Keskitalo}, R.},
  \bibinfo{author}{{Kisner}, T.S.}, \bibinfo{author}{{Kneissl}, R.},
  \bibinfo{author}{{Knoche}, J.}, \bibinfo{author}{{Kunz}, M.},
  \bibinfo{author}{{Kurki-Suonio}, H.}, \bibinfo{author}{{Lagache}, G.},
  \bibinfo{author}{{L{\"a}hteenm{\"a}ki}, A.}, \bibinfo{author}{{Lamarre},
  J.M.}, \bibinfo{author}{{Lasenby}, A.}, \bibinfo{author}{{Lawrence}, C.R.},
  \bibinfo{author}{{Leonardi}, R.}, \bibinfo{author}{{Levrier}, F.},
  \bibinfo{author}{{Liguori}, M.}, \bibinfo{author}{{Lilje}, P.B.},
  \bibinfo{author}{{Linden-V{\o}rnle}, M.},
  \bibinfo{author}{{L{\'o}pez-Caniego}, M.}, \bibinfo{author}{{Lubin}, P.M.},
  \bibinfo{author}{{Mac{\'\i}as-P{\'e}rez}, J.F.}, \bibinfo{author}{{Maffei},
  B.}, \bibinfo{author}{{Magalh{\~a}es}, A.M.}, \bibinfo{author}{{Maino}, D.},
  \bibinfo{author}{{Mandolesi}, N.}, \bibinfo{author}{{Maris}, M.},
  \bibinfo{author}{{Marshall}, D.J.}, \bibinfo{author}{{Martin}, P.G.},
  \bibinfo{author}{{Mart{\'\i}nez-Gonz{\'a}lez}, E.}, \bibinfo{author}{{Masi},
  S.}, \bibinfo{author}{{Matarrese}, S.}, \bibinfo{author}{{Mazzotta}, P.},
  \bibinfo{author}{{Melchiorri}, A.}, \bibinfo{author}{{Mendes}, L.},
  \bibinfo{author}{{Mennella}, A.}, \bibinfo{author}{{Migliaccio}, M.},
  \bibinfo{author}{{Miville-Desch{\^e}nes}, M.A.}, \bibinfo{author}{{Moneti},
  A.}, \bibinfo{author}{{Montier}, L.}, \bibinfo{author}{{Morgante}, G.},
  \bibinfo{author}{{Mortlock}, D.}, \bibinfo{author}{{Munshi}, D.},
  \bibinfo{author}{{Murphy}, J.A.}, \bibinfo{author}{{Naselsky}, P.},
  \bibinfo{author}{{Nati}, F.}, \bibinfo{author}{{Natoli}, P.},
  \bibinfo{author}{{Netterfield}, C.B.}, \bibinfo{author}{{Noviello}, F.},
  \bibinfo{author}{{Novikov}, D.}, \bibinfo{author}{{Novikov}, I.},
  \bibinfo{author}{{Oxborrow}, C.A.}, \bibinfo{author}{{Pagano}, L.},
  \bibinfo{author}{{Pajot}, F.}, \bibinfo{author}{{Paladini}, R.},
  \bibinfo{author}{{Paoletti}, D.}, \bibinfo{author}{{Pasian}, F.},
  \bibinfo{author}{{Perdereau}, O.}, \bibinfo{author}{{Perotto}, L.},
  \bibinfo{author}{{Perrotta}, F.}, \bibinfo{author}{{Piacentini}, F.},
  \bibinfo{author}{{Piat}, M.}, \bibinfo{author}{{Pietrobon}, D.},
  \bibinfo{author}{{Plaszczynski}, S.}, \bibinfo{author}{{Poidevin}, F.},
  \bibinfo{author}{{Pointecouteau}, E.}, \bibinfo{author}{{Polenta}, G.},
  \bibinfo{author}{{Popa}, L.}, \bibinfo{author}{{Pratt}, G.W.},
  \bibinfo{author}{{Prunet}, S.}, \bibinfo{author}{{Puget}, J.L.},
  \bibinfo{author}{{Rachen}, J.P.}, \bibinfo{author}{{Reach}, W.T.},
  \bibinfo{author}{{Rebolo}, R.}, \bibinfo{author}{{Reinecke}, M.},
  \bibinfo{author}{{Remazeilles}, M.}, \bibinfo{author}{{Renault}, C.},
  \bibinfo{author}{{Ricciardi}, S.}, \bibinfo{author}{{Riller}, T.},
  \bibinfo{author}{{Ristorcelli}, I.}, \bibinfo{author}{{Rocha}, G.},
  \bibinfo{author}{{Rosset}, C.}, \bibinfo{author}{{Roudier}, G.},
  \bibinfo{author}{{Rusholme}, B.}, \bibinfo{author}{{Sandri}, M.},
  \bibinfo{author}{{Savini}, G.}, \bibinfo{author}{{Scott}, D.},
  \bibinfo{author}{{Spencer}, L.D.}, \bibinfo{author}{{Stolyarov}, V.},
  \bibinfo{author}{{Stompor}, R.}, \bibinfo{author}{{Sudiwala}, R.},
  \bibinfo{author}{{Sutton}, D.}, \bibinfo{author}{{Suur-Uski}, A.S.},
  \bibinfo{author}{{Sygnet}, J.F.}, \bibinfo{author}{{Tauber}, J.A.},
  \bibinfo{author}{{Terenzi}, L.}, \bibinfo{author}{{Toffolatti}, L.},
  \bibinfo{author}{{Tomasi}, M.}, \bibinfo{author}{{Tristram}, M.},
  \bibinfo{author}{{Tucci}, M.}, \bibinfo{author}{{Umana}, G.},
  \bibinfo{author}{{Valenziano}, L.}, \bibinfo{author}{{Valiviita}, J.},
  \bibinfo{author}{{Van Tent}, B.}, \bibinfo{author}{{Vielva}, P.},
  \bibinfo{author}{{Villa}, F.}, \bibinfo{author}{{Wade}, L.A.},
  \bibinfo{author}{{Wandelt}, B.D.}, \bibinfo{author}{{Zonca}, A.},
  \bibinfo{year}{2015}.
\newblock \bibinfo{title}{Planck intermediate results. xxi. comparison of
  polarized thermal emission from galactic dust at 353 ghz with interstellar
  polarization in the visible}.
\newblock \bibinfo{journal}{Astronomy \& Astrophysics} \bibinfo{volume}{576},
  \bibinfo{pages}{A106}.
\newblock \DOIprefix\doi{10.1051/0004-6361/201424087},
  \href{http://arxiv.org/abs/1405.0873}{\tt arXiv:1405.0873}.
%Type = Article
\bibitem[{{Planck Collaboration} et~al.(2016){Planck Collaboration}, {Ade},
  {Aghanim}, {Alves}, {Arnaud}, {Arzoumanian}, {Ashdown}, {Aumont},
  {Baccigalupi}, {Band ay}, {Barreiro}, {Bartolo}, {Battaner}, {Benabed},
  {Beno{\^\i}t}, {Benoit-L{\'e}vy}, {Bernard}, {Bersanelli}, {Bielewicz},
  {Bock}, {Bonavera}, {Bond}, {Borrill}, {Bouchet}, {Boulanger}, {Bracco},
  {Burigana}, {Calabrese}, {Cardoso}, {Catalano}, {Chiang}, {Christensen},
  {Colombo}, {Combet}, {Couchot}, {Crill}, {Curto}, {Cuttaia}, {Danese},
  {Davies}, {Davis}, {de Bernardis}, {de Rosa}, {de Zotti}, {Delabrouille},
  {Dickinson}, {Diego}, {Dole}, {Donzelli}, {Dor{\'e}}, {Douspis}, {Ducout},
  {Dupac}, {Efstathiou}, {Elsner}, {En{\ss}lin}, {Eriksen},
  {Falceta-Gon{\c{c}}alves}, {Falgarone}, {Ferri{\`e}re}, {Finelli}, {Forni},
  {Frailis}, {Fraisse}, {Franceschi}, {Frejsel}, {Galeotta}, {Galli}, {Ganga},
  {Ghosh}, {Giard}, {Gjerl{\o}w}, {Gonz{\'a}lez-Nuevo}, {G{\'o}rski},
  {Gregorio}, {Gruppuso}, {Gudmundsson}, {Guillet}, {Harrison}, {Helou},
  {Hennebelle}, {Henrot-Versill{\'e}}, {Hern{\'a}ndez-Monteagudo}, {Herranz},
  {Hildebrand t}, {Hivon}, {Holmes}, {Hornstrup}, {Huffenberger}, {Hurier},
  {Jaffe}, {Jaffe}, {Jones}, {Juvela}, {Keih{\"a}nen}, {Keskitalo}, {Kisner},
  {Knoche}, {Kunz}, {Kurki-Suonio}, {Lagache}, {Lamarre}, {Lasenby},
  {Lattanzi}, {Lawrence}, {Leonardi}, {Levrier}, {Liguori}, {Lilje},
  {Linden-V{\o}rnle}, {L{\'o}pez-Caniego}, {Lubin}, {Mac{\'\i}as-P{\'e}rez},
  {Maino}, {Mandolesi}, {Mangilli}, {Maris}, {Martin},
  {Mart{\'\i}nez-Gonz{\'a}lez}, {Masi}, {Matarrese}, {Melchiorri}, {Mendes},
  {Mennella}, {Migliaccio}, {Miville-Desch{\^e}nes}, {Moneti}, {Montier},
  {Morgante}, {Mortlock}, {Munshi}, {Murphy}, {Naselsky}, {Nati},
  {Netterfield}, {Noviello}, {Novikov}, {Novikov}, {Oppermann}, {Oxborrow},
  {Pagano}, {Pajot}, {Paladini}, {Paoletti}, {Pasian}, {Perotto}, {Pettorino},
  {Piacentini}, {Piat}, {Pierpaoli}, {Pietrobon}, {Plaszczynski},
  {Pointecouteau}, {Polenta}, {Ponthieu}, {Pratt}, {Prunet}, {Puget}, {Rachen},
  {Reinecke}, {Remazeilles}, {Renault}, {Renzi}, {Ristorcelli}, {Rocha},
  {Rossetti}, {Roudier}, {Rubi{\~n}o-Mart{\'\i}n}, {Rusholme}, {Sandri},
  {Santos}, {Savelainen}, {Savini}, {Scott}, {Soler}, {Stolyarov}, {Sudiwala},
  {Sutton}, {Suur-Uski}, {Sygnet}, {Tauber}, {Terenzi}, {Toffolatti}, {Tomasi},
  {Tristram}, {Tucci}, {Umana}, {Valenziano}, {Valiviita}, {Van Tent},
  {Vielva}, {Villa}, {Wade}, {Wandelt}, {Wehus}, {Ysard}, {Yvon} and
  {Zonca}}]{planck2016}
\bibinfo{author}{{Planck Collaboration}}, \bibinfo{author}{{Ade}, P.A.R.},
  \bibinfo{author}{{Aghanim}, N.}, \bibinfo{author}{{Alves}, M.I.R.},
  \bibinfo{author}{{Arnaud}, M.}, \bibinfo{author}{{Arzoumanian}, D.},
  \bibinfo{author}{{Ashdown}, M.}, \bibinfo{author}{{Aumont}, J.},
  \bibinfo{author}{{Baccigalupi}, C.}, \bibinfo{author}{{Band ay}, A.J.},
  \bibinfo{author}{{Barreiro}, R.B.}, \bibinfo{author}{{Bartolo}, N.},
  \bibinfo{author}{{Battaner}, E.}, \bibinfo{author}{{Benabed}, K.},
  \bibinfo{author}{{Beno{\^\i}t}, A.}, \bibinfo{author}{{Benoit-L{\'e}vy}, A.},
  \bibinfo{author}{{Bernard}, J.P.}, \bibinfo{author}{{Bersanelli}, M.},
  \bibinfo{author}{{Bielewicz}, P.}, \bibinfo{author}{{Bock}, J.J.},
  \bibinfo{author}{{Bonavera}, L.}, \bibinfo{author}{{Bond}, J.R.},
  \bibinfo{author}{{Borrill}, J.}, \bibinfo{author}{{Bouchet}, F.R.},
  \bibinfo{author}{{Boulanger}, F.}, \bibinfo{author}{{Bracco}, A.},
  \bibinfo{author}{{Burigana}, C.}, \bibinfo{author}{{Calabrese}, E.},
  \bibinfo{author}{{Cardoso}, J.F.}, \bibinfo{author}{{Catalano}, A.},
  \bibinfo{author}{{Chiang}, H.C.}, \bibinfo{author}{{Christensen}, P.R.},
  \bibinfo{author}{{Colombo}, L.P.L.}, \bibinfo{author}{{Combet}, C.},
  \bibinfo{author}{{Couchot}, F.}, \bibinfo{author}{{Crill}, B.P.},
  \bibinfo{author}{{Curto}, A.}, \bibinfo{author}{{Cuttaia}, F.},
  \bibinfo{author}{{Danese}, L.}, \bibinfo{author}{{Davies}, R.D.},
  \bibinfo{author}{{Davis}, R.J.}, \bibinfo{author}{{de Bernardis}, P.},
  \bibinfo{author}{{de Rosa}, A.}, \bibinfo{author}{{de Zotti}, G.},
  \bibinfo{author}{{Delabrouille}, J.}, \bibinfo{author}{{Dickinson}, C.},
  \bibinfo{author}{{Diego}, J.M.}, \bibinfo{author}{{Dole}, H.},
  \bibinfo{author}{{Donzelli}, S.}, \bibinfo{author}{{Dor{\'e}}, O.},
  \bibinfo{author}{{Douspis}, M.}, \bibinfo{author}{{Ducout}, A.},
  \bibinfo{author}{{Dupac}, X.}, \bibinfo{author}{{Efstathiou}, G.},
  \bibinfo{author}{{Elsner}, F.}, \bibinfo{author}{{En{\ss}lin}, T.A.},
  \bibinfo{author}{{Eriksen}, H.K.},
  \bibinfo{author}{{Falceta-Gon{\c{c}}alves}, D.},
  \bibinfo{author}{{Falgarone}, E.}, \bibinfo{author}{{Ferri{\`e}re}, K.},
  \bibinfo{author}{{Finelli}, F.}, \bibinfo{author}{{Forni}, O.},
  \bibinfo{author}{{Frailis}, M.}, \bibinfo{author}{{Fraisse}, A.A.},
  \bibinfo{author}{{Franceschi}, E.}, \bibinfo{author}{{Frejsel}, A.},
  \bibinfo{author}{{Galeotta}, S.}, \bibinfo{author}{{Galli}, S.},
  \bibinfo{author}{{Ganga}, K.}, \bibinfo{author}{{Ghosh}, T.},
  \bibinfo{author}{{Giard}, M.}, \bibinfo{author}{{Gjerl{\o}w}, E.},
  \bibinfo{author}{{Gonz{\'a}lez-Nuevo}, J.}, \bibinfo{author}{{G{\'o}rski},
  K.M.}, \bibinfo{author}{{Gregorio}, A.}, \bibinfo{author}{{Gruppuso}, A.},
  \bibinfo{author}{{Gudmundsson}, J.E.}, \bibinfo{author}{{Guillet}, V.},
  \bibinfo{author}{{Harrison}, D.L.}, \bibinfo{author}{{Helou}, G.},
  \bibinfo{author}{{Hennebelle}, P.}, \bibinfo{author}{{Henrot-Versill{\'e}},
  S.}, \bibinfo{author}{{Hern{\'a}ndez-Monteagudo}, C.},
  \bibinfo{author}{{Herranz}, D.}, \bibinfo{author}{{Hildebrand t}, S.R.},
  \bibinfo{author}{{Hivon}, E.}, \bibinfo{author}{{Holmes}, W.A.},
  \bibinfo{author}{{Hornstrup}, A.}, \bibinfo{author}{{Huffenberger}, K.M.},
  \bibinfo{author}{{Hurier}, G.}, \bibinfo{author}{{Jaffe}, A.H.},
  \bibinfo{author}{{Jaffe}, T.R.}, \bibinfo{author}{{Jones}, W.C.},
  \bibinfo{author}{{Juvela}, M.}, \bibinfo{author}{{Keih{\"a}nen}, E.},
  \bibinfo{author}{{Keskitalo}, R.}, \bibinfo{author}{{Kisner}, T.S.},
  \bibinfo{author}{{Knoche}, J.}, \bibinfo{author}{{Kunz}, M.},
  \bibinfo{author}{{Kurki-Suonio}, H.}, \bibinfo{author}{{Lagache}, G.},
  \bibinfo{author}{{Lamarre}, J.M.}, \bibinfo{author}{{Lasenby}, A.},
  \bibinfo{author}{{Lattanzi}, M.}, \bibinfo{author}{{Lawrence}, C.R.},
  \bibinfo{author}{{Leonardi}, R.}, \bibinfo{author}{{Levrier}, F.},
  \bibinfo{author}{{Liguori}, M.}, \bibinfo{author}{{Lilje}, P.B.},
  \bibinfo{author}{{Linden-V{\o}rnle}, M.},
  \bibinfo{author}{{L{\'o}pez-Caniego}, M.}, \bibinfo{author}{{Lubin}, P.M.},
  \bibinfo{author}{{Mac{\'\i}as-P{\'e}rez}, J.F.}, \bibinfo{author}{{Maino},
  D.}, \bibinfo{author}{{Mandolesi}, N.}, \bibinfo{author}{{Mangilli}, A.},
  \bibinfo{author}{{Maris}, M.}, \bibinfo{author}{{Martin}, P.G.},
  \bibinfo{author}{{Mart{\'\i}nez-Gonz{\'a}lez}, E.}, \bibinfo{author}{{Masi},
  S.}, \bibinfo{author}{{Matarrese}, S.}, \bibinfo{author}{{Melchiorri}, A.},
  \bibinfo{author}{{Mendes}, L.}, \bibinfo{author}{{Mennella}, A.},
  \bibinfo{author}{{Migliaccio}, M.}, \bibinfo{author}{{Miville-Desch{\^e}nes},
  M.A.}, \bibinfo{author}{{Moneti}, A.}, \bibinfo{author}{{Montier}, L.},
  \bibinfo{author}{{Morgante}, G.}, \bibinfo{author}{{Mortlock}, D.},
  \bibinfo{author}{{Munshi}, D.}, \bibinfo{author}{{Murphy}, J.A.},
  \bibinfo{author}{{Naselsky}, P.}, \bibinfo{author}{{Nati}, F.},
  \bibinfo{author}{{Netterfield}, C.B.}, \bibinfo{author}{{Noviello}, F.},
  \bibinfo{author}{{Novikov}, D.}, \bibinfo{author}{{Novikov}, I.},
  \bibinfo{author}{{Oppermann}, N.}, \bibinfo{author}{{Oxborrow}, C.A.},
  \bibinfo{author}{{Pagano}, L.}, \bibinfo{author}{{Pajot}, F.},
  \bibinfo{author}{{Paladini}, R.}, \bibinfo{author}{{Paoletti}, D.},
  \bibinfo{author}{{Pasian}, F.}, \bibinfo{author}{{Perotto}, L.},
  \bibinfo{author}{{Pettorino}, V.}, \bibinfo{author}{{Piacentini}, F.},
  \bibinfo{author}{{Piat}, M.}, \bibinfo{author}{{Pierpaoli}, E.},
  \bibinfo{author}{{Pietrobon}, D.}, \bibinfo{author}{{Plaszczynski}, S.},
  \bibinfo{author}{{Pointecouteau}, E.}, \bibinfo{author}{{Polenta}, G.},
  \bibinfo{author}{{Ponthieu}, N.}, \bibinfo{author}{{Pratt}, G.W.},
  \bibinfo{author}{{Prunet}, S.}, \bibinfo{author}{{Puget}, J.L.},
  \bibinfo{author}{{Rachen}, J.P.}, \bibinfo{author}{{Reinecke}, M.},
  \bibinfo{author}{{Remazeilles}, M.}, \bibinfo{author}{{Renault}, C.},
  \bibinfo{author}{{Renzi}, A.}, \bibinfo{author}{{Ristorcelli}, I.},
  \bibinfo{author}{{Rocha}, G.}, \bibinfo{author}{{Rossetti}, M.},
  \bibinfo{author}{{Roudier}, G.}, \bibinfo{author}{{Rubi{\~n}o-Mart{\'\i}n},
  J.A.}, \bibinfo{author}{{Rusholme}, B.}, \bibinfo{author}{{Sandri}, M.},
  \bibinfo{author}{{Santos}, D.}, \bibinfo{author}{{Savelainen}, M.},
  \bibinfo{author}{{Savini}, G.}, \bibinfo{author}{{Scott}, D.},
  \bibinfo{author}{{Soler}, J.D.}, \bibinfo{author}{{Stolyarov}, V.},
  \bibinfo{author}{{Sudiwala}, R.}, \bibinfo{author}{{Sutton}, D.},
  \bibinfo{author}{{Suur-Uski}, A.S.}, \bibinfo{author}{{Sygnet}, J.F.},
  \bibinfo{author}{{Tauber}, J.A.}, \bibinfo{author}{{Terenzi}, L.},
  \bibinfo{author}{{Toffolatti}, L.}, \bibinfo{author}{{Tomasi}, M.},
  \bibinfo{author}{{Tristram}, M.}, \bibinfo{author}{{Tucci}, M.},
  \bibinfo{author}{{Umana}, G.}, \bibinfo{author}{{Valenziano}, L.},
  \bibinfo{author}{{Valiviita}, J.}, \bibinfo{author}{{Van Tent}, B.},
  \bibinfo{author}{{Vielva}, P.}, \bibinfo{author}{{Villa}, F.},
  \bibinfo{author}{{Wade}, L.A.}, \bibinfo{author}{{Wandelt}, B.D.},
  \bibinfo{author}{{Wehus}, I.K.}, \bibinfo{author}{{Ysard}, N.},
  \bibinfo{author}{{Yvon}, D.}, \bibinfo{author}{{Zonca}, A.},
  \bibinfo{year}{2016}.
\newblock \bibinfo{title}{Planck intermediate results. xxxv. probing the role
  of the magnetic field in the formation of structure in molecular clouds}.
\newblock \bibinfo{journal}{Astronomy \& Astrophysics} \bibinfo{volume}{586},
  \bibinfo{pages}{A138}.
\newblock \DOIprefix\doi{10.1051/0004-6361/201525896},
  \href{http://arxiv.org/abs/1502.04123}{\tt arXiv:1502.04123}.
%Type = Article
\bibitem[{{Schleuning}(1998)}]{Schleuning1998}
\bibinfo{author}{{Schleuning}, D.A.}, \bibinfo{year}{1998}.
\newblock \bibinfo{title}{Far-infrared and submillimeter polarization of omc-1:
  Evidence for magnetically regulated star formation}.
\newblock \bibinfo{journal}{Astrophysical Journal} \bibinfo{volume}{493},
  \bibinfo{pages}{811--825}.
\newblock \DOIprefix\doi{10.1086/305139}.
%Type = Article
\bibitem[{{Shu} et~al.(1987){Shu}, {Adams} and {Lizano}}]{shu87}
\bibinfo{author}{{Shu}, F.H.}, \bibinfo{author}{{Adams}, F.C.},
  \bibinfo{author}{{Lizano}, S.}, \bibinfo{year}{1987}.
\newblock \bibinfo{title}{Star formation in molecular clouds: observation and
  theory.}
\newblock \bibinfo{journal}{Annual Review of Astronomy and Astrophysics}
  \bibinfo{volume}{25}, \bibinfo{pages}{23--81}.
\newblock \DOIprefix\doi{10.1146/annurev.aa.25.090187.000323}.
%Type = Inproceedings
\bibitem[{{Shu} et~al.(1999){Shu}, {Allen}, {Shang}, {Ostriker} and
  {Li}}]{shu99}
\bibinfo{author}{{Shu}, F.H.}, \bibinfo{author}{{Allen}, A.},
  \bibinfo{author}{{Shang}, H.}, \bibinfo{author}{{Ostriker}, E.C.},
  \bibinfo{author}{{Li}, Z.Y.}, \bibinfo{year}{1999}.
\newblock \bibinfo{title}{Low-mass star formation: Theory}, in:
  \bibinfo{editor}{{Lada}, C.J.}, \bibinfo{editor}{{Kylafis}, N.D.} (Eds.),
  \bibinfo{booktitle}{NATO Advanced Science Institutes (ASI) Series C},
  \bibinfo{publisher}{Kluwer Academic Publishers}. p. \bibinfo{pages}{193}.
%Type = Article
\bibitem[{{Suriano} et~al.(2019){Suriano}, {Li}, {Krasnopolsky}, {Suzuki} and
  {Shang}}]{Surianoetal2019MNRAS}
\bibinfo{author}{{Suriano}, S.S.}, \bibinfo{author}{{Li}, Z.Y.},
  \bibinfo{author}{{Krasnopolsky}, R.}, \bibinfo{author}{{Suzuki}, T.K.},
  \bibinfo{author}{{Shang}, H.}, \bibinfo{year}{2019}.
\newblock \bibinfo{title}{The formation of rings and gaps in wind-launching
  non-ideal mhd discs: three-dimensional simulations}.
\newblock \bibinfo{journal}{\mnras} \bibinfo{volume}{484},
  \bibinfo{pages}{107--124}.
\newblock \DOIprefix\doi{10.1093/mnras/sty3502},
  \href{http://arxiv.org/abs/1810.02234}{\tt arXiv:1810.02234}.
%Type = Article
\bibitem[{{Tomisaka}(2011)}]{tomisaka11}
\bibinfo{author}{{Tomisaka}, K.}, \bibinfo{year}{2011}.
\newblock \bibinfo{title}{{Origin of Molecular Outflow Determined from Thermal
  Dust Polarization}}.
\newblock \bibinfo{journal}{\pasj} \bibinfo{volume}{63}, \bibinfo{pages}{147}.
\newblock \DOIprefix\doi{10.1093/pasj/63.1.147},
  \href{http://arxiv.org/abs/1011.0125}{\tt arXiv:1011.0125}.
%Type = Article
\bibitem[{{Wurster} and {Li}(2018)}]{wurster18}
\bibinfo{author}{{Wurster}, J.}, \bibinfo{author}{{Li}, Z.Y.},
  \bibinfo{year}{2018}.
\newblock \bibinfo{title}{The role of magnetic fields in the formation of
  protostellar discs}.
\newblock \bibinfo{journal}{Frontiers in Astronomy and Space Sciences}
  \bibinfo{volume}{5}, \bibinfo{pages}{39}.
\newblock \DOIprefix\doi{10.3389/fspas.2018.00039},
  \href{http://arxiv.org/abs/1812.06728}{\tt arXiv:1812.06728}.

\end{thebibliography}
\bibliographystyle{model2-names.bst}\biboptions{authoryear}
\end{document}